\documentclass[10pt,conference]{IEEEtran}

\makeatletter
\def\ps@headings{%
\def\@oddhead{\mbox{}\scriptsize\rightmark \hfil \thepage}%
\def\@evenhead{\scriptsize\thepage \hfil \leftmark\mbox{}}%
\def\@oddfoot{}%
\def\@evenfoot{}}
\makeatother 

\IEEEoverridecommandlockouts
\usepackage{bbm}
\usepackage{amsfonts}
\usepackage[dvips]{graphicx}
\usepackage{times}
\usepackage{cite}
\usepackage{amsmath}
\usepackage{array}
\usepackage{amssymb}

\usepackage{stfloats}
\usepackage{slashbox}
\usepackage{graphicx}
\usepackage{footnote}
\usepackage{amsthm}
\usepackage{booktabs}
\usepackage{array}
\usepackage{algorithmic}
\usepackage{algorithm}
\usepackage{subeqnarray}
\usepackage{cases}
\usepackage{threeparttable}
\usepackage{color}
\usepackage{epstopdf}
\usepackage{bm}
\usepackage{subcaption}
\usepackage{dsfont}

\usepackage[labelformat=simple]{subcaption}

\begin{document}

\title{Content-Centric Multicast Beamforming in Cache-Enabled Cloud Radio Access Networks}
\author{\IEEEauthorblockN{Hao Zhou$^*$, Meixia Tao$^*$, Erkai Chen$^*$, Wei Yu$^\dag$}
	\IEEEauthorblockA{*Dept. of Electronic Engineering, Shanghai Jiao Tong University, Shanghai, China \\
		 $\dag$Dept. of Electrical and Computer Engineering, University of Toronto, Toronto, Canada \\
		 	Email: $\lbrace$zhouhao$\_$zh, mxtao, cek1006$\rbrace$@sjtu.edu.cn, weiyu@comm.utoronto.ca}}
\maketitle

\begin{abstract}
Multicast transmission and wireless caching are effective ways of reducing air and backhaul traffic load in wireless networks. This paper proposes to incorporate these two key ideas for content-centric multicast transmission in a cloud radio access network (RAN) where multiple base stations (BSs) are connected to a central processor (CP) via finite-capacity backhaul links. Each BS has a cache with finite storage size and is equipped with multiple antennas. The BSs cooperatively transmit contents, which are either stored in the local cache or fetched from the CP, to multiple users in the network. Users requesting a same content form a multicast group and are served by a same cluster of BSs cooperatively using multicast beamforming. Assuming fixed cache placement, this paper investigates the joint design of multicast beamforming and content-centric BS clustering by formulating an optimization problem of minimizing the total network cost under the quality-of-service (QoS) constraints for each multicast group. The network cost involves both the transmission power and the backhaul cost. We model the backhaul cost using the mixed $\ell_0/\ell_2$-norm of beamforming vectors. To solve this non-convex problem, we first approximate it using the semidefinite relaxation (SDR) method and concave smooth functions. We then propose a difference of convex functions (DC) programming algorithm to obtain suboptimal solutions and show the connection of three smooth functions. Simulation results validate the advantage of multicasting and show the effects of different cache size and caching policies in cloud RAN.

\end{abstract}
\section{Introduction} 
Cloud radio access network (RAN) is an emerging network architecture capable of exploiting the advantage of multicell cooperation in the future fifth-generation (5G) wireless system\cite{6898939}. In a cloud RAN, the base stations (BSs) are connected to a central processor (CP) via digital backhaul links, thus enabling joint data processing and precoding capabilities across multiple BSs. This paper proposes a {\em content-centric} view for cloud RAN design. We equip the BSs with finite-size cache, where popular contents desired by multiple users can be stored. We formulate and solve a network optimization problem while accounting for the finite-capacity backhaul links between the BSs and the CP. 

To address the issue of limited backhaul, previous works on wireless cooperative networks \cite{zhao2013coordinated,dai2013sparse,zhuang2014backhaul} consider the problem of minimizing the backhaul traffic and transmission power by designing sparse beamformer and user-centric BS clustering. Further, \cite{dai2014sparse} considers the weighted sum rate (WSR) optimization problem under per-BS backhaul constraints. However, all these works focus on the unicast scenario and promote a user-centric view of system design without considering the effect of caching.

Recently, {\em wireless caching} has been investigated as an effective way of reducing peak traffic and backhaul load. By deploying caches at BSs and placing popular contents in them in advance, the issue of limited backhaul capacity can be addressed fundamentally. In \cite{6495773}, the authors show that with small or even no backhaul capacity, femto-caching can support high demand of wireless video distribution. In \cite{6763007}, the upper and lower bounds of the capacity of the caching system are derived, and it is shown that the network capacity could be further improved by using coded multicasting for content delivery. These studies motivate us to consider the {\em cache-enabled} cloud RAN, where each BS is equipped with a cache with finite storage size. Compared with cooperative networks without caching, cache-enabled cloud RAN can fundamentally reduce the backhaul cost and support more flexible BS clustering. We note that in \cite{junzhang}, a similar wireless caching network has been considered, where the authors study the data assignment and unicast beamforming design. 

Different from previous work focused on unicast \cite{zhao2013coordinated,dai2013sparse,zhuang2014backhaul,junzhang}, where data is transmitted to each user individually no matter whether the actual contents requested by different users are the same or not, this paper focuses on the problem of multicast transmission. We assume that multiple users can request the same content, and the content is delivered using multicast beamforming to these users on the same resource block. Compared with traditional unicast, multicast can improve energy and spectral efficiency. In addition, since the popular contents cached in the BSs are possibly requested by multiple users, multicast could better exploit the potential of wireless caching. 

This paper studies the joint design of multicast beamforming and content-centric BS clustering, which differs from the fixed BS clustering in coordinated muticell multicast networks \cite{6340379} or user-centric BS clustering in unicast systems\cite{dai2013sparse}. In each scheduling interval, the BS clustering is dynamically optimized with respect to each multicast group. We formulate an optimization problem with the objective of minimizing the total power consumption as well as the backhaul cost under the quality-of-service (QoS) constraints for each multicast group. The backhaul cost is formulated as a function of the mixed $\ell_0/\ell_2$-norm of the beamforming vectors. The challenge in solving such a problem is due to both the non-convex QoS constraints and the $\ell_0$-norm in the backhaul cost. In this paper, we first use the semidefinite relaxation (SDR) method introduced in \cite{karipidis2008quality} to handle the non-convex QoS constraints. We then adopt the smooth function approach in sparse signal processing to approximate the $\ell_0$-norm with concave smooth functions.

In sparse signal processing, one approach to handle the $\ell_0$-norm minimization problem is to approximate the $\ell_0$-norm with its reweighted $\ell_1$-norm \cite{candes2008enhancing} and update the weight factors iteratively. Another approach is the smooth function method \cite{zhuang2014backhaul}, where the authors use Gaussian family functions to approximate the $\ell_0$-norm. The smooth function method is a better approximation to the $\ell_0$-norm but its performance highly depends on the smoothness factor of the approximation function. In this paper, we adopt the smooth function approach and solve the approximated problem with a difference of convex functions (DC) algorithm \cite{horst1999dc}. We explore the use of three different smooth functions, the logarithmic function, the exponential function, and the arctangent function, and show that with a particular weight updating rule, all three are equivalent to the reweighted $\ell_1$-norm minimization\cite{candes2008enhancing}. Simulation results are presented to illustrate the performance of proposed algorithm and the benefit of wireless caching. 

\emph{Notations}: Boldface uppercase letters denote matrices and boldface lowercase letters denote column vectors. The sets of complex numbers and binary numbers are denoted as $\mathbb{C}$ and $\mathbb{B}$ respectively. The statistical expectation, transpose and Hermitian transpose are denoted as $\mathbb{E}(\cdot)$, $(\cdot)^{T}$ and $(\cdot)^{H}$ respectively. The Frobenius norm and the $\ell_0$-norm are denoted as $\|\cdot\|_2$ and $\|\cdot\|_0$ respectively. An all-one vector of length $M$ is denoted as $\bm{1}_M$. An all-zero vector of length $M$ is denoted as $\bm{0}_M$. The inner product of matrices $\bm{X}$ and $\bm{Y}$ is defined as $\left\langle \bm{X}, \bm{Y}\right\rangle = \text{Tr}(\bm{X}^H\bm{Y})$. For a square matrix $\bm{S}_{M \times M}$, $\bm{S}\succeq \bm{0}$ means that $\bm{S}$ is positive semidefinite.

\section{System Model}
\subsection{Signal Model}
We consider a cache-enabled cloud RAN with one CP, $L$ BSs and $K$ mobile users. Each BS is equipped with $N_t$ antennas and each user has a single antenna. Scheduling and beamformer design are done at the CP. Each BS is connected to the CP via a finite-capacity backhaul link. The total number of contents is $F$; different contents are independent. The CP stores all the contents and there is a cache at each BS, which stores finite number of contents. Each user requests a content according to the content popularity, and users requesting the same content form a multicast group. We assume that the total number of multicast groups is $M$. The set of users in group $m$ is denoted as $\mathcal{K}_m$ with $\sum_{m=1}^M |\mathcal{K}_m|=K$. 

\begin{figure}[!t]
	\begin{center}
		\includegraphics*[width=0.45\textwidth]{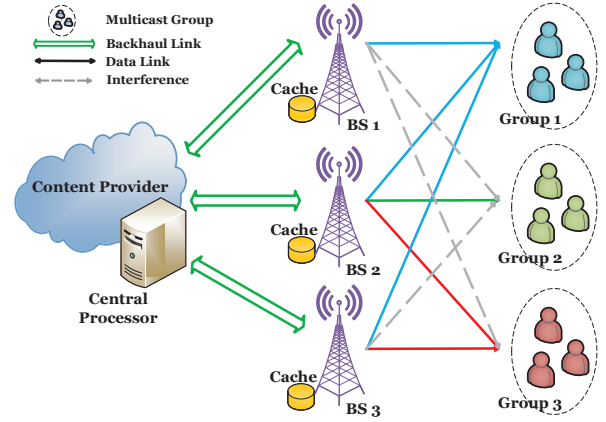}
	\end{center}
	\vspace{-0.2cm}
	\caption{An example of downlink cloud RAN with $M=3$ groups and $L=3$ cache-enabled BSs connected to a CP via digital backhaul links, where each multicast group is served by a cluster of BSs.} \label{fig:systemModel}
	\vspace{-0.2cm}
\end{figure}

We study the cooperative downlink multicast transmission and dynamic content-centric BS clustering. Each group $m$ is served by a cluster of BSs cooperatively, denoted as $\mathcal{Q}_m$. In each scheduling interval, the BS clustering $\{\mathcal{Q}_m \}_{m=1}^M$ is dynamically optimized by the CP. For example, in Fig.~\ref{fig:systemModel}, the instantaneous BS clusters for different groups are $\mathcal{Q}_1 = \{1,2,3\}$, $\mathcal{Q}_2 = \{2\}$ and $\mathcal{Q}_3 = \{2,3\}$, respectively. For BS $3$, since it serves both group $1$  and group $3$, it should acquire the contents for these two groups either from its local cache or through backhaul. 

We denote the aggregate beamforming vector of group $m$ from all BSs as $\bm{w}_m \in \mathbb{C}^{LN_t\times 1} = [{\bm{w}^H_{1,m}}, {\bm{w}^H_{2,m}}, \cdots, {\bm{w}^H_{L,m}}]^H$, where $\bm{w}_{l,m}\in \mathbb{C}^{N_t\times 1}$ is the beamforming vector for group $m$ at BS $l$. Note that the BS clustering is implicitly defined by the beamforming vectors. If the beamforming vector $\bm{w}_{l,m}$ is $\bm{0}_{N_t}$, then BS $l$ does not serve group $m$ and is thus not in $\mathcal{Q}_m$. On the other hand, if $\bm{w}_{l,m} \neq \bm{0}_{N_t}$, BS $l$ is part of the serving cluster of group $m$. Thus, the size of the BS cooperation cluster for group $m$ can be expressed as a mixed $\ell_0/\ell_2$-norm of the beamforming vector $\{\bm{w}_{l,m}\}_{l=1}^{L}$, i.e. $ | \mathcal{Q}_m | = \sum_{l=1}^L \big \| \|\bm{w}_{l,m} \|^2_2 \big \|_0$. 

We denote the data symbol of the content requested by group $m$ as $s_m \in \mathbb{C}$, with $\mathbb{E}\left[|s_m|^2\right]=1$. For user $k\in \mathcal{K}_m$, its received downlink signal $y_k$ can be written as
\begin{equation}
y_k = \bm{h}_{k}^H\bm{w}_{m}s_m + \sum_{n\neq m}^M\bm{h}_{k}^H\bm{w}_{n}s_{n} + z_k,
\end{equation}
where $\bm{h}_{k} \in \mathbb{C}^{LN_t\times 1}$ is the network-wide channel vector from all BSs to user $k$ and $z_k \sim \mathcal{CN}(0,\sigma^2)$ is the addictive noise.

The received signal-to-interference-plus-noise ratio (SINR) at user $k\in \mathcal{K}_m$ is
\begin{equation}
\text{SINR}_{k} = \frac{| \bm{h}_{k}^H\bm{w}_{m}|^2}{\sum_{n\neq m}^M|\bm{h}_{k}^H\bm{w}_{n}|^2 + \sigma^2}.
\end{equation}

We define the target SINR vector as $\bm{\gamma} = [\gamma_1,\gamma_2,\cdots,\gamma_M]$ with each element $\gamma_m$ being the target SINR to be achieved by the users in group $m$. In this paper, we consider the fixed rate transmission as in \cite{dai2013sparse}, where the transmission rate for group $m$ is set as $R_m= \log_2(1+\gamma_m)$. Thus, to successfully decode the message, for any user $k \in \mathcal{K}_m$, its achievable data rate should be larger than $R_m$, that is, for $\forall m, k \in \mathcal{K}_m$, $\log_2(1+\text{SINR}_k) \geqslant R_m$. 

\subsection{Cache Model}
We assume that each content has normalized size of $1$ and the local storage size of BS $l$ is $F_l\left(F_l< F\right)$, which is also the maximum number of contents it can store. Therefore, we define a cache placement matrix $\bm{C} \in \mathbb{B}^{L\times F}$, where $c_{l,f}=1$ means the content $f$ is cached in BS $l$ and $c_{l,f}=0$ means the opposite. Note that $\forall l$, $\sum_{f=1}^F c_{l,f}\leqslant F_l$.

We assume that the cache placement is static, that is, matrix $\bm{C}$ is fixed and is known at the CP (similar assumptions have been made in previous literature, e.g., \cite{junzhang}). This assumption is based on the fact that the optimization of cache placement is performed on a large time scale; while the beamforming design is done on a small time scale of channel coherence time. Hence, it is reasonable to assume that during a short scheduling interval, the cache placement policy remains unchanged. 

\subsection{Cost Model}
We consider the total network cost which consists of both the transmission power and the backhaul cost. Let $f_m$ denote the content requested by users in multicast group $m$. For BS $l$ in $\mathcal{Q}_m$, if content $f_m$ is in its cache, it can access the content directly without costing backhaul. On the contrary, if content $f_m$ is not cached, BS $l$ needs to first fetch this content from the CP via the backhaul link. Since the data rates of fetching the contents from the CP need to be as large as the content-delivery rate, the backhaul cost in this case is modeled as the transmission rates of multicast groups. 

The total backhaul cost at all BSs can be written as
\begin{equation} \label{equ:backhaul}
C_{B} = \sum_{m=1}^M \sum_{l=1}^L \big \| \|\bm{w}_{l,m} \|^2_2 \big \|_0 R_m (1-c_{l,f_m}).
\end{equation} 

The total network cost can be written as
\begin{equation}
C_N = \eta\underbrace{\sum_{m=1}^M \|\bm{w}_{m} \|^2_2}_{\text{Power Consumption}} + \underbrace{\sum_{m=1}^M\sum_{l=1}^L \big \| \|\bm{w}_{l,m} \|^2_2 \big \|_0 R_m (1-c_{l,f_m})}_{\text{Backhaul Cost}},
\end{equation}
where $\eta$ is a weight parameter. By adjusting the value of $\eta$, we can emphasize on one cost versus the other.

Note that in a network without caching, there is a tradeoff between power and backhaul cost. To reduce power consumption, each group can be served by more BSs, which increases backhaul cost. However, in cache-enabled cloud RAN, for each group, the BSs caching the requested content can be involved in the cooperative cluster of the group without costing extra backhaul.

\section{Problem Formulation and Approximation}
In this section, we present the optimization problem of minimizing the total network cost by jointly designing multicast beamforming and BS clustering. We show that this problem is a non-convex problem and further approximate it with two steps.

\subsection{Problem Formulation}
Our objective is to minimize the total network cost, under the constraints of the peak transmission power at each BS and the SINR requirement of each multicast group. 

The optimization problem is formulated as
\begin{subequations} \label{problem0}
\begin{flalign}
\mathcal{P}_0: \quad \underset{\{ \bm{w}_m\}_{m=1}^M}{\text{minimize}} & \quad C_N   \\
\text{subject to} & \quad \text{SINR}_{k} \geqslant \gamma_{m}, \forall m, k \in \mathcal{K}_m \\
& \quad \sum_{m=1}^M \|\bm{w}_{l,m} \|^2_2 \leqslant P_l, \forall l
\end{flalign}
\end{subequations}
where $P_l$ is the peak transmission power at BS $l$. 

\begin{figure}[!t]
	\begin{center}
		\includegraphics*[width=0.45\textwidth]{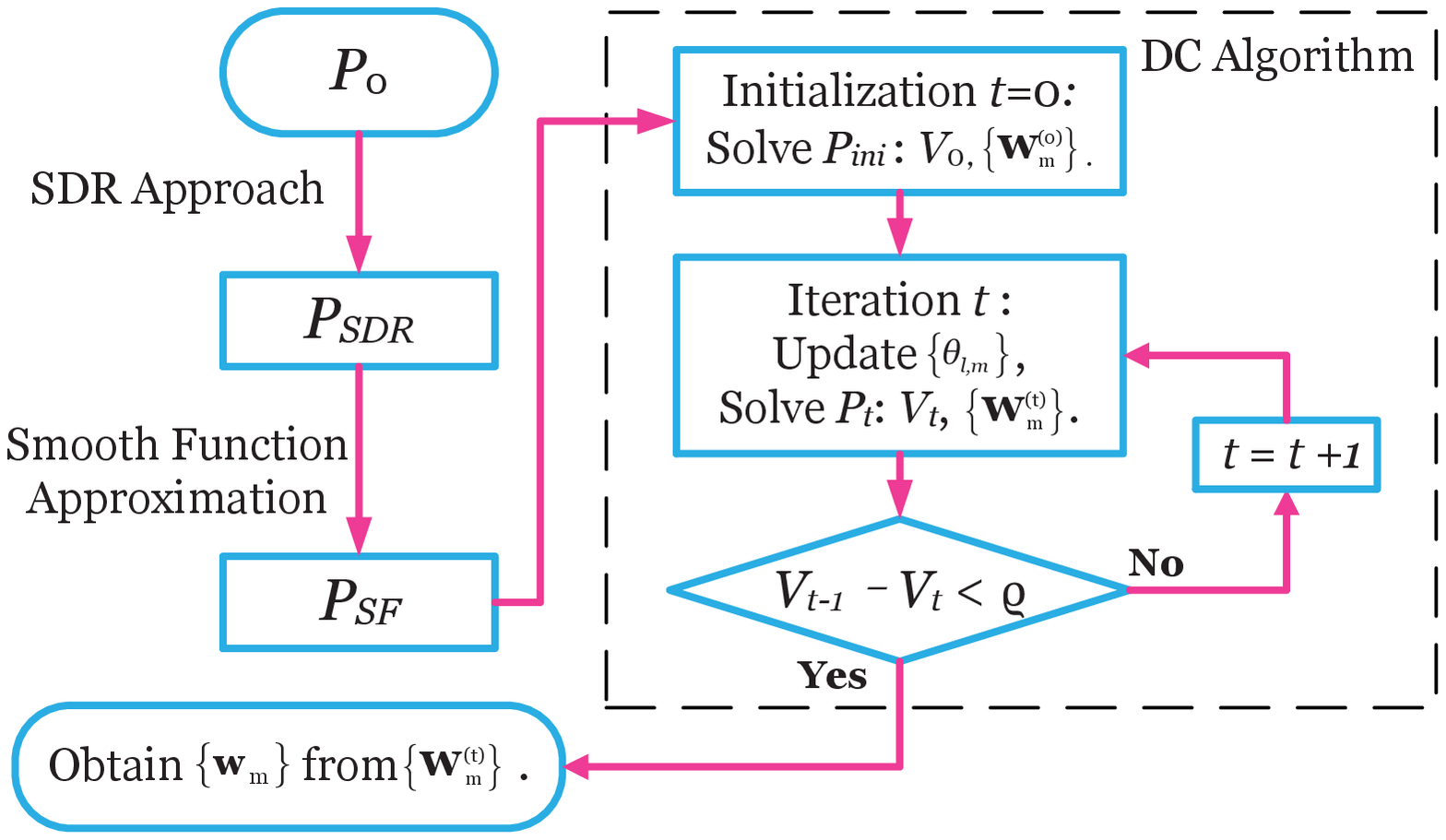}
	\end{center}
	\vspace{-0.4cm}
	\caption{Overall procedure for solving problem $\mathcal{P}_0$} \label{fig:liucheng}
	\vspace{-0.2cm}
\end{figure}

Problem $\mathcal{P}_0$ is a non-convex problem, where the non-convexity comes from both the $\ell_0$-norm in the objective function and the SINR constraints. Unlike traditional unicast beamforming problem where the non-convex SINR constraints can be transformed to a second-order cone programming (SOCP) problem and the optimal solutions can be obtained with convex optimization, the multicast beamforming problem is NP-hard \cite{karipidis2008quality}. In this paper, we use two techniques to approximate problem $\mathcal{P}_0$, namely {\em SDR relaxation} and {\em $\ell_0$-norm approximation}. The overall procedure to solve $\mathcal{P}_0$ is shown in Fig.~\ref{fig:liucheng}, with each step elaborated in following sections. 

\subsection{Step 1 -- SDR Relaxation}
In both single-cell \cite{karipidis2008quality} and multicell \cite{6340379} scenarios, the semidefinite relaxation (SDR) method has been used to deal with the non-convex SINR constraints in multicast beamforming design problems. 

We define two sets of matrices $\{\bm{W}_m\in\mathbb{C}^{LN_t\times LN_t} \}_{m=1}^M$ and $\{\bm{H}_k\in\mathbb{C}^{LN_t\times LN_t}\}_{k=1}^K$ as
\begin{equation}
\bm{W}_m = \bm{w}_{m}\bm{w}_{m}^H \quad \text{and}\quad \bm{H}_k = \bm{h}_{k}\bm{h}_{k}^H, \quad \forall m,k.
\end{equation}

We further define a set of selecting matrices $\{\bm{J}_l\}_{l=1}^L$, where each matrix $\bm{J}_l\in\mathbb{B}^{LN_t\times LN_t}$ is a diagonal matrix defined as
\begin{equation}
\bm{J}_l = \text{diag}\left(\left[ \bm{0}_{(l-1)N_t}^H, \bm{1}_{N_t}^H, \bm{0}_{(L-l)N_t}^H\right]\right),\quad \forall l.
\end{equation}

Therefore, we have 
\begin{equation}
\|\bm{w}_{l,m}\|^2_2 = \text{Tr}(\bm{W}_{m}\bm{J}_l), \forall l,m. 
\end{equation}

By adopting the SDR method, problem $\mathcal{P}_0$ can be relaxed and rewritten as
\begin{subequations} \label{problem_sdr}
	\begin{align}
		& \mathcal{P}_{SDR: } \notag \\
		& \underset{\{ \bm{W}_m\}_{m=1}^M}{\text{minimize}} && \sum_{m=1}^M \eta\text{Tr}\left(\bm{W}_{m}\right) + \sum_{m=1}^M\sum_{l=1}^L\alpha_{l,m}\|\text{Tr}\left(\bm{W}_{m}\bm{J}_l \right)\|_0   \\
		&\text{subject to} && \frac{\text{Tr}(\bm{W}_{m}\bm{H}_{k})}{\sum_{n\neq m}^M\text{Tr}(\bm{W}_{n}\bm{H}_{k})+\sigma^2} \geqslant \gamma_{m}, \forall m, k \in \mathcal{K}_m \label{c1} \\
		&&& \sum_{m=1}^M \text{Tr}(\bm{W}_{m}\bm{J}_l) \leqslant P_l, \quad \forall l \label{c2} \\
		&&& {\bm{W}_{m} \succeq \bm{0}, \quad \forall m \label{c3}}
	\end{align}
\end{subequations}
where, to further simplify the mathematical representation, we have defined a set of constants $\{\alpha_{l,m}\}_{l=1, \cdots, L}^{m=1, \cdots, M}$ with $\alpha_{l,m} = R_m\left(1-c_{l,f_m}\right)$. 

The SINR constraints in problem $\mathcal{P}_{SDR}$ are convex. We denote the resulting optimal $\{\bm{W}_m\}$ after solving problem $\mathcal{P}_{SDR}$ as $\{\bm{W}^*_m\}$. If $\bm{W}^*_{m}$ is already rank-one, then for group $m$ the optimal aggregate beamformer $\bm{w}^*_{m}$ of problem $\mathcal{P}_0$ can be obtained by applying eigen-value decomposition to $\bm{W}^*_{m}$ as $\bm{W}^*_{m} = \lambda^*_{m}\hat{\bm{w}}_{m}{\hat{\bm{w}}^H_{m}}$ and taking $\bm{w}^*_{m} = \sqrt{\lambda^*_{m}}\hat{\bm{w}}_{m}$. Otherwise, the beamforming vectors $\{\bm{w}_{m} \}$ can be generated with the {\em randomization} method used in \cite{6340379} and \cite{karipidis2008quality}. 

\subsection{Step 2 -- $\ell_0$-norm Approximation}
To solve the problem $\mathcal{P}_{SDR}$, we further approximate the non-convex $\ell_0$-norm in the objective with a continuous function denoted as $f(x)$. Specifically, we consider three frequently used smooth functions: logarithmic function, exponential function and arctangent function \cite{smoothFunction}, defined as
\begin{equation} \label{equ:sf}
f(\bm{X}) = 
\begin{cases}
\log\left(\frac{\text{Tr}(\bm{X})+\theta}{\theta}\right), & \text{for log-function} \\
1-e^{-\frac{\text{Tr}(\bm{X})}{\theta}}, & \text{for exp-function} \\
\frac{2}{\pi}\arctan\left({\frac{\text{Tr}(\bm{X})}{\theta}}\right), & \text{for atan-function}
\end{cases}
\end{equation}
where $\theta$ is a parameter to adjust the smoothness of the functions. In all three cases, with larger $\theta$, the function is smoother but is a worse approximation to the $\ell_0$-norm.

In \cite{zhuang2014backhaul}, the authors use the Gaussian family smooth functions, where the $\ell_0$-norm of $\|\bm{w}\|_2$ is approximated with $f(\|\bm{w}\|_2) = 1-\text{exp}(\frac{-\|\bm{w}\|_2^2}{2\theta^2})$. In this work, by adopting the exponential smooth function, the $\ell_0$-norm is approximated with $f_{exp}(\bm{W}) = 1-\text{exp}(\frac{-\text{Tr}(\bm{W})}{\theta})$. Comparing $f(\|\bm{w}\|_2)$ and $f_{exp}(\bm{W})$, we can see that the exponential smooth function in \eqref{equ:sf} has the same approximation effect as the Gaussian smooth function in \cite{zhuang2014backhaul}.

With smooth function, $\mathcal{P}_{SDR}$ can be rewritten as
\begin{subequations} \label{problem_appx}
	\begin{flalign}
	&\mathcal{P}_{SF}: \notag \\
	 &\underset{\{ \bm{W}_m\}_{m=1}^M}{\text{minimize}} && \sum_{m=1}^M \eta\text{Tr}\left(\bm{W}_{m}\right) + \sum_{m=1}^M\sum_{l=1}^L\alpha_{l,m}f\left(\bm{W}_{m}\bm{J}_l\right)  \\
	&\text{subject to} &&\eqref{c1},\eqref{c2}, \eqref{c3}.
	\end{flalign}
\end{subequations}

For ease of presentation, we express the objective as the summation of two functions $G(\bm{W})$ and $F(\bm{W})$, defined as
\begin{equation}
G(\bm{W}) = \sum_{m=1}^M \eta\text{Tr}\left(\bm{W}_{m}\right) \text{ and }
F(\bm{W}) = \sum_{m=1}^M\sum_{l=1}^L\alpha_{l,m}f_{l,m},
\end{equation}
where $f_{l,m} = f\left(\bm{W}_{m}\bm{J}_l\right),  \forall l,m$.

We see that $G(\bm{W})$ and $F(\bm{W})$ are an affine and a concave function of $\bm{W}$, respectively, so problem $\mathcal{P}_{SF}$ can be viewed as the difference of two continuous convex functions with convex constraints. Therefore, this problem can be solved with the DC algorithm, which falls in the category of majorization minimization (MM) algorithms \cite{horst1999dc}. 

\section{DC Based Algorithm and Analysis}
In this section, we first present the DC based algorithm for solving the problem $\mathcal{P}_{SF}$ using the logarithmic smooth function. We then show that the resulting algorithm can also be interpreted as a DC algorithm with the other two smooth functions and a different $\theta$ updating rule. 

\subsection{DC Based Algorithm with Log-Function}
The DC algorithm iteratively optimizes an approximated convex function of the original concave objective function and produces a sequence of improving $\{\bm{W}_m\}$. The algorithm converges to some global/local optimal solution. 

The initial $\{ \bm{W}^{(0)}_m\}$ is found by solving the following power minimization problem
\begin{subequations} \label{problem_ini}
	\begin{align}
	& \mathcal{P}_{ini} : \quad V_{ini} \triangleq && \underset{\{ \bm{W}_m\}_{m=1}^M}{\text{minimize}} && \sum_{m=1}^M \text{Tr}\left(\bm{W}_{m}\right) \\
	&&&\text{subject to} && \eqref{c1},\eqref{c2}, \eqref{c3}.
	\end{align}
\end{subequations}

In the $t$-th iteration, $\{ \bm{W}_m^{(t)}\}$ is generated as the solution of the approximated convex optimization problem,
\begin{subequations} \label{problem_it}
	\begin{align} \label{pt}
	&\mathcal{P}_t: && V_{t} \ \triangleq \ \underset{\{ \bm{W}_m\}_{m=1}^M}{\text{minimize}} \quad G(\bm{W})  \notag\\
	&&&  + \sum_{m=1}^M\sum_{l=1}^L\alpha_{l,m}\left\langle \left(\nabla_{\bm{W}_m^{(t-1)}} f_{l,m}\right), \left(\bm{W}_m-\bm{W}_m^{(t-1)}\right)\right\rangle    \\	
	&&& \quad \quad \quad \text{subject to} \quad \eqref{c1}, \eqref{c2}, \eqref{c3}.
	\end{align}
\end{subequations}
where $\nabla_{\bm{W}_m^{(t-1)}} f_{l,m}$ is the gradient matrix of $f_{l,m}$ at $\bm{W}_m^{(t-1)}$. 

Specifically, for log-function, the gradient matrix $\nabla_{\bm{W}_m} f_{l,m} \in \mathbb{C}^{LN_t\times LN_t}$ at $\{\bm{W}_m^{(t)} \}$ is
\begin{equation}  \label{equ:log2}
\bm{D}_{log}(l,m) = \nabla_{\bm{W}_m^{(t)}} f_{l,m} = \frac{\bm{J}_l}{\text{Tr}\left(\bm{W}_m^{(t)}\bm{J}_l\right)+\theta_{l,m}},
\end{equation}

If we let the smoothness factor $\theta_{l,m} = \epsilon$, where $\epsilon$ is a very small positive constant. Then we have
\begin{equation} \label{equ:log}
\bm{D}_{log}(l,m) = \frac{\bm{J}_l}{\text{Tr}\left(\bm{W}_m^{(t)}\bm{J}_l\right) + \epsilon}.
\end{equation} 

Note that \eqref{equ:log} has the similar form as the weight factor of the reweighted $\ell_1$-norm approach in \cite{candes2008enhancing}. Thus, the DC algorithm with log-function and smoothness factor $\theta_{l,m}\rightarrow 0$ is just the reweighted $\ell_1$-norm minimization algorithm of \cite{dai2013sparse}.

In \eqref{pt}, function $F(\bm{W})$ is approximated with its first-order Taylor expansion, which provides an upper bound. This algorithm terminates when the sequence of $\{ \bm{W}^{(t)}_m\}$ converges to some stationary point, and the objective value $V_t$ converges, that is, $V_{t-1}-V_t<\varrho$, where $\varrho$ is a small constant. 

\subsection{Updating Rule of $\theta$ for Other Smooth Functions}
The performance of $\ell_0$-norm approximation algorithms depends on the smoothness factor $\theta$. Intuitively, when $x$ is large, $\theta$ should be large so that the approximation algorithm can explore the entire parameter space; when $x$ is small, $\theta$ should be small so that $f(x)$ has behavior close to $\ell_0$-norm. In \cite{4663911}, the authors propose to use a decreasing sequence of $\theta$, but the updating rule does not depend on $x$.

In this paper, we explore a novel $\theta$ updating rule that achieves the above effect automatically using a sequence of $\theta$ which depends on specific $x$ in each iteration. More specifically, we propose to set $\theta$ to be the one that maximizes the gradient of the approximation function.

Note that the gradient matrices of the exponential and arctangent functions in \eqref{equ:sf} are, respectively,
\begin{equation} \label{equ:exp_dif}
\bm{D}_{exp}(l,m)  = \frac{\bm{J}_l}{\theta_{l,m}}e^{-\frac{\text{Tr}\left(\bm{W}_{m}\bm{J}_l\right)}{\theta_{l,m}}},
\end{equation}
and
\begin{equation} \label{equ:atan_dif}
\bm{D}_{atan}(l,m) = \frac{2}{\pi}\cdot\frac{\bm{J}_l}{\theta_{l,m}\left(\frac{\text{Tr}\left(\bm{W}_{m}\bm{J}_l\right)}{\theta_{l,m}}\right)^2+\theta_{l,m}}.
\end{equation}

An interesting observation is that for all three functions in \eqref{equ:sf}, if we maximize their gradients, we get expressions of the same form. Specifically, for the log-function, we get \eqref{equ:log} with optimal $\theta^*_{l,m} \rightarrow 0$. For the exp-function and atan-function, we get 
\begin{equation} \label{equ:exp}
\bm{D}_{exp}^*(l,m) = \max_{\theta_{l,m}} \bm{D}_{exp}(l,m) = \frac{\bm{J}_l}{e \cdot \text{Tr}(\bm{W}_{m}^{(t)}\bm{J}_l)},
\end{equation}

\begin{equation} \label{equ_atan}
\bm{D}_{atan}^*(l,m) = \max_{\theta_{l,m}} \bm{D}_{atan}(l,m) = \frac{\bm{J}_l}{\pi \cdot \text{Tr}(\bm{W}_{m}^{(t)}\bm{J}_l) },
\end{equation}
respectively with the optimal $\theta^*_{l,m} = \text{Tr}(\bm{W}_{m}^{(t)}\bm{J}_l)$.

We see that the gradient matrices for all three approximation functions in \eqref{equ:sf} differ by a constant multiple only. Therefore, they lead to the same algorithm if we update $\theta_{l,m}$ such that $\nabla_{\bm{W}_m^{(t)}} f_{l,m}$ is maximized in the $t$-th iteration, i.e.,
\begin{equation} \label{equ:theta}
\theta^*_{l,m} \triangleq \arg\max_{\theta_{l,m}} \nabla_{\bm{W}_m^{(t)}} f_{l,m}.
\end{equation}

In this proposed algorithm, the approximation functions are adjusted dynamically to achieve a good tradeoff between smoothness and approximation to $\ell_0$-norm. Further, similarity between \eqref{equ:log}, \eqref{equ:exp}, and \eqref{equ_atan} means that with proposed $\theta$ updating rule, these three approximation functions lead to almost the same performance. 

\section{Simulation Results}
We consider a cache-enabled cloud RAN covering an area of circle with the radius of $1.2$km, where 7 BSs ($L=7, N_t=3$) are placed in a equilateral triangular lattice with the distance between adjacent BSs of $0.8k$m. The total number of contents is $F=100$. A total number of $140$ users are distributed in this network with uniform distribution and they are scheduled in a round-robin manner. In each scheduling interval, $K=14$ users are scheduled. We assume half of the scheduled users request a common content (e.g., a live video) and each of the rest randomly requests one content according to the content popularity, which is modeled as Zipf distribution with skewness parameter 1. Users requesting the same content participate in the same multicast group. We assume all BSs have the same cache size. The channels between BSs and users are generated with a normalized Rayleigh fading component and a distance-dependent path loss, modeled as $PL$(dB)$ = 148.1 + 37.6\log_{10}(d)$ with $8$dB log-normal shadowing, where $d$ is the distance from the user to the BS. The transmit antenna power gain at each BS is 10 dBi. The power spectral density of downlink noise is $-172$dBm/Hz with the channel bandwidth of $10$MHz. The peak transmission power of each BS is $P_l = 10$W$, \forall l$. The target SINR of each content is $10$dB. We set $\varrho = 10^{-6}$ for the convergence condition in the DC algorithm and $\epsilon = 10^{-7}$ in \eqref{equ:log}. Each simulation result is averaged over $300$ scheduling intervals.

\begin{figure}[t]
	\begin{center}
		\includegraphics[scale=.34]{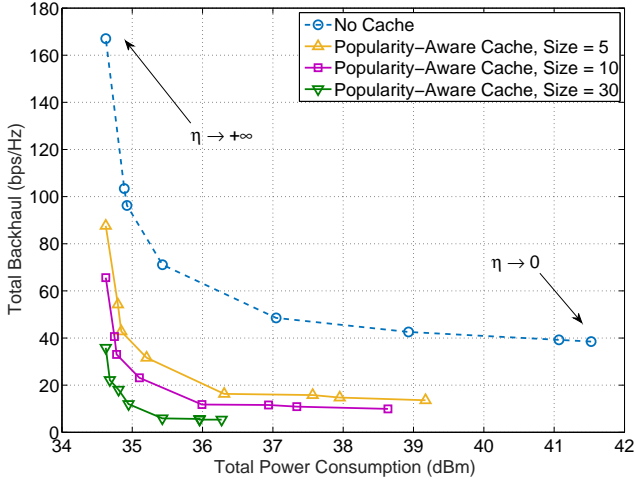}
	\end{center}
	\vspace{-0.3em}
	\caption{\small{Power-backhaul tradeoff for different cache size.}} \label{fig:cacheSize}
	\vspace{-0.2em}
\end{figure}

In Fig.~\ref{fig:cacheSize}, we show the effects of wireless caching and the cache size. The popularity-aware cache refers to the policy where each BS caches the contents with highest popularity. The figure shows that compared with the network without cache, the cache-enabled network can reduce the backhaul cost by more than $50\%$ when each BS only caches $5\%$  of the total contents. The backhaul reduction is up to $75\%$ when each BS can cache $30\%$ of the total contents. 

\begin{figure}[t]
	\begin{center}
		\includegraphics[scale=.34]{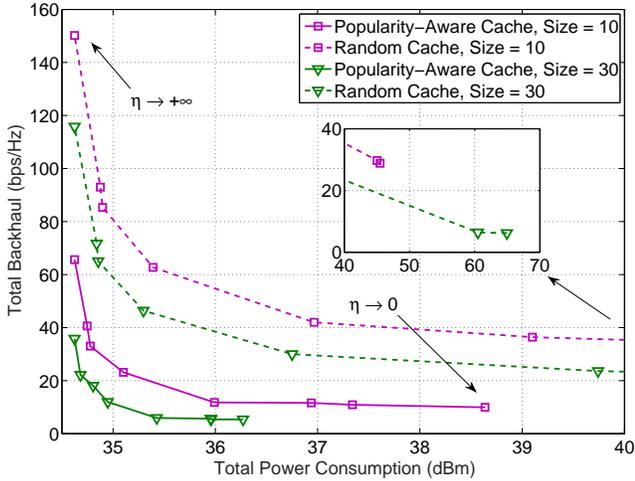}
	\end{center}
	\vspace{-0.3em}
	\caption{\small{Power-backhaul tradeoff for different cache policies.}} \label{fig:alg}
	\vspace{-1.2em}
\end{figure}

In Fig.~\ref{fig:alg}, we compare the effects of different cache strategies. In the random cache policy, the contents are randomly cached with equal probability in each BS. The result shows that the popularity-aware cache has better power-backhaul efficiency in general. In specific, with the same transmission power cost of 38dBm, and the same cache size of 10, popularity-aware cache only costs about $1/4$ backhaul cost of the random cache policy. However, in the extreme case when we do not consider the total transmission power, the minimum backhaul costs of the two caching strategies are about the same if the cache size is 30.

\begin{figure}[t]
	\begin{center}
		\includegraphics[scale=.34]{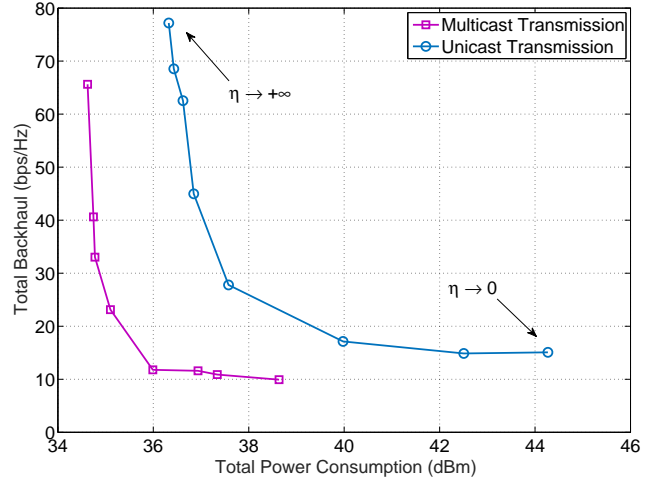}
	\end{center}
	\vspace{-0.3em}
	\caption{\small{Power-backhaul tradeoff for multicast and unicast transmission.}} \label{fig:mu}
	 \vspace{-1.5em}
\end{figure}

In Fig.~\ref{fig:mu}, we compare the power-backhaul tradeoff of multicast and unicast transmission in the same scenario. We use popularity-aware caching policy with the cache size of $10$. In unicast transmission, users of the same group are served by different beamformers and we use the algorithm proposed in \cite{dai2013sparse}, where in each iteration, for $\forall k,l$, the weight factor $\rho_k^l$ is set to $0$ if the content requested by user $k$ is cached in BS $l$. If multiple users in the same group are served by a same BS, the backhaul cost is counted only once. The figure shows that in the power-limited system, the total power consumption of multicast transmission is about $2$dB less than the unicast scenario. When the total power consumption is $38$dBm, the backhaul cost of multicast transmission is only $1/3$ of unicast transmission. These observations validates the advantage of caching and multicasting in such a scenario.

\section{Conclusion}
This paper investigates the joint design of multicast beamforming and content-centric BS clustering in a cache-enabled cloud RAN. The optimization problem is formulated as the minimization of the total network cost, including the power consumption and the backhaul cost, under the QoS constraint of each multicast group. We adopt the SDR method and the smooth function approach, introduced in sparse signal processing, to approximate this non-convex problem as a DC programming problem. We then propose a DC algorithm to obtain sub-optimal solutions. Further, we propose a new smoothness updating method and give insight into its connection to reweighted $\ell_1$-norm minimization. Simulation results show that, compared with unicast transmission, multicast transmission can achieve better power-backhaul tradeoff, and the backhaul cost can be further reduced with larger cache size.

\bibliographystyle{IEEEtran}
\bibliography{IEEEabrv,main} 

\end{document}